\documentstyle[aps,prb,epsfig]{revtex}
\begin{document}
\renewcommand{\thefigure}{\arabic{figure}}
\baselineskip=0.5cm
\title{Self-consistent Overhauser model for the pair distribution function of an electron gas in dimensionalities $D=3$ and $D=2$}
\author{B. Davoudi$^{1,2}$, M. Polini$^1$, R. Asgari$^{1,2}$ and M. P. Tosi$^1$}
\address{$^1$NEST-INFM and Classe di Scienze, Scuola Normale Superiore, I-56126 Pisa, Italy\\
$^2$Institute for Studies in Theoretical Physics and Mathematics, Tehran
19395-5531, Iran\\}
\maketitle
\vspace{0.2 cm}

\begin{abstract}
We present self-consistent calculations of the spin-averaged pair distribution function $g(r)$ 
for a homogeneous electron gas in the paramagnetic state in both three and two dimensions, based on an extension of a model that was originally proposed by A. W. Overhauser [Can. J. Phys. {\bf 73}, 683 (1995)] 
and further evaluated by P. Gori-Giorgi and J. P. Perdew [Phys. Rev. B {\bf 64}, 155102 (2001)]. 
The model involves the solution of a two-electron scattering problem via an effective Coulombic potential, 
that we determine within a self-consistent Hartree approximation. We find numerical results for $g(r)$
that are in excellent agreement with Quantum Monte Carlo data at low and intermediate coupling strength $r_s$, 
extending up to $r_s\approx 10$ in dimensionality $D=3$. However, the Hartree approximation
does not properly account for the emergence of a first-neighbor peak at stronger coupling, such
as at $r_s=5$ in $D=2$, and has limited accuracy in regard to the spin-resolved components $g_{\uparrow\uparrow}(r)$ and $g_{\uparrow\downarrow}(r)$. We also report calculations of the electron-electron s-wave scattering length, to test an analytical expression proposed by Overhauser in $D=3$ and to present new results in $D=2$ at moderate coupling strength. Finally, we indicate how this approach can be 
extended to evaluate the pair distribution functions in inhomogeneous electron systems and hence to obtain 
improved exchange-correlation energy functionals.
\end{abstract}
\vspace{0.5cm}


\hspace{2.4cm} 71.45.Gm - Exchange, correlation, dielectric and magnetic functions, plasmons
\section{Introduction}
Many of the electron-electron interaction effects in simple metals and semiconductors can be understood by
reference to the homogeneous electron-gas model. A central role is played by the electron-pair distribution
function $g(r)$, which describes how short-range exchange and correlations enter to determine the probability
of finding two electrons at a relative distance $r$.\cite{review} The mean potential energy of the electron gas can
be calculated from $g(r)$ and hence, through an integration over its dependence on the coupling strength, the total
energy as well. Thus, an accurate knowledge of this function is crucial for applications of density functional theory \cite{Hohenberg} (DFT) in various schemes that
have been proposed to transcend the local density approximation (LDA) in the construction of exchange and correlation energy
functionals.\cite{Gunnarsson}

A precise definition of $g(r)$ is through the average number of electrons contained in a spherical shell of radius $r$ and thickness $dr$ centered on an electron at the origin, which is given by $ng(r)\Omega_Dr^{D-1}dr$ where $D$ is the space dimensionality, $n$ the electron density, and $\Omega_D$ the solid angle in  $D$ dimensions
(with $\Omega_2=2\pi$ and $\Omega_3=4\pi$). In fact, $g(r)$ is the average of the distribution functions for parallel- and antiparallel-spin electron pairs, $g(r)=[g_{\uparrow\uparrow}(r)+g_{\uparrow\downarrow}(r)]/2$. While this spin-average reflects the charge-charge correlations in the electron gas, the spin-spin correlations are described by the difference distribution function $g_d(r)=[g_{\uparrow\uparrow}(r)-g_{\uparrow\downarrow}(r)]/2$.

Early calculations of the pair distribution functions in the $3D$ electron gas were based on the use of a Bijl-Jastrow correlated wave function for the ground state\cite{???} and on exploiting the fluctuation-dissipation theorem for a self-consistent determination from the charge and spin response functions.\cite{Singwi} These early results were validated with the advent of the Quantum Monte Carlo (QMC) techniques,\cite{Ceperley} which have produced a wealth of accurate data on correlation and response functions over a wide range of coupling strength for both the $3D$ 
\cite{Ceperley,Ortiz} and the $2D$ \cite{Tanatar} case. The QMC data have in turn stimulated a number of further theoretical studies.

Here we are specifically concerned with the approach proposed by Overhauser \cite{Overhauser2} for the evaluation of the
value $g(0)$ of the pair distribution function at contact. While $g_{\uparrow\uparrow}(0)$ vanishes on account of the Pauli principle, $g_{\uparrow\downarrow}(0)$ is determined by the two-body scattering events and in Overhauser's model was obtained from the solution of an effective Schr\"odinger equation for the relative motion of two electrons with
antiparallel spins. Earlier work had demonstrated an exact cusp condition \cite{Kimball} relating the logarithmic
derivative of $g_{\uparrow\downarrow}(r)$ at contact to the Bohr radius, and had emphasized the importance of the
electron-electron ladder diagrams\cite{Yasuhara} in evaluating $g_{\uparrow\downarrow}(0)$. The approach of Overhauser
has subsequently been used to evaluate $g_{\uparrow\downarrow}(0)$ in the $2D$ electron gas,\cite{Polini} and has been extended by Gori-Giorgi and Perdew \cite{gp} to evaluate $g(r)$ at finite $r$ in $3D$ through an accurate numerical solution of Overhauser's two-body Schr\"odinger equation. Their results are in good agreement with QMC data in
the short-range part of $g(r)$.

In the present work we further develop this approach to the evaluation of the pair distribution functions by
(i) adopting a self-consistent Hartree scheme for the determination of the effective potential entering the two-body Schr\"odinger equation; and (ii) carrying out calculations for both a $3D$ and a $2D$ electron gas. The paper is organized as follows. Section \ref{sec2} presents the essential theoretical background and Section \ref{sec3} reports and discusses our numerical results. In Section \ref{sec4} we indicate how this approach could be extended to evaluate pair correlations in an inhomogeneous electron gas and hence to obtain improved exchange-correlation energy functionals for applications of DFT. A brief summary concludes the paper in Section \ref{sec4}.

\section{Essential theory}\label{sec2}
Following the work of Overhauser\cite{Overhauser2} and of Gori-Giorgi and Perdew\cite{gp}, we aim to solve the 
two-electron scattering problem in some effective interaction potential $V(r)$ (assumed to be independent of the spin
state of the electron pair) in order to determine the wave functions 
$\Psi^{\uparrow\uparrow}_{{\bf k},{\bf K}}({\bf r},{\bf R})$ and
$\Psi^{\uparrow\downarrow}_{{\bf k},{\bf K}}({\bf r},{\bf R})$ for the parallel and antiparallel spin states. Here, 
${\bf r}={\bf r}_1-{\bf r}_2$ and ${\bf R}=({\bf r}_1+{\bf r}_2)/2$ are the relative coordinate and the center-of-mass
coordinate of the pair, the conjugate momenta being ${\bf k}$ and ${\bf K}$. The spin-resolved pair distribution  functions $g_{\sigma \sigma'}(r)$ can then be obtained as
\begin{equation}\label{eq7}
g_{\sigma \sigma'}(r)=\langle\langle |\Psi_{{\bf k}, {\bf K}}^{\sigma \sigma'}({\bf r}, {\bf R})|^2\rangle_{\theta} \rangle_{p(k)}
\end{equation}
where $\langle\langle...\rangle_{\theta} \rangle_{p(k)}$ represents averages over the scattering angle and over the probability $p(k)$ of finding two electrons with relative momentum $k$ in the electron gas,
\begin{equation}
\langle\langle{\mathcal O}\rangle_{\theta} \rangle_{p(k)}=(1/\Omega_{D})\int_0^{k_F} dk\, p(k)\int d \Omega_{D}\,{\mathcal O}.
\end{equation}
The function $p(k)$ can be calculated from the momentum distribution $n(k)$ \cite{gp,perdew}, using
\begin{equation}\label{pofk}
p(k)=\frac{\Omega_D\,k^{D-1}}{n^2\,(2 \pi)^D}\,\int \,\frac{d^{D}{\bf q}}{(2\pi)^D}\, n(q)n(|{\bf q} +{\bf k}|)\, .
\end{equation}
The prefactor in front of the integral is obtained from the condition that $p(k)$ should integrate to unity.

The pair wave functions can be written in the form of angular-momentum expansions,
\begin{equation}\label{eq2}
\Psi_{{\bf k}, {\bf K}}^{\uparrow \uparrow}({\bf r}, {\bf R})=\sqrt{2}\, \frac{\exp{(i\, {\bf K}\cdot{\bf R})}}{r^{(D-1)/2}}\,\sum_{\ell=1 \,\,({\mathrm odd})}^{\infty}i^{\ell}\,{\mathcal A}^{(D)}_{\ell}(\theta)\,\Phi^{(D)}_{\ell, {\bf k}}(r)
\end{equation}
and
\begin{equation}\label{eq3}
\Psi_{{\bf k}, {\bf K}}^{\uparrow \downarrow}({\bf r}, {\bf R})=\frac{\exp{(i\, {\bf K}\cdot{\bf R})}}{r^{(D-1)/2}}\,\left
[\Phi^{(D)}_{0, \,{\bf k}}(r)+\sum_{\ell=1}^{\infty}i^{\ell}\,{\mathcal A}^{(D)}_{\ell}(\theta)\,\Phi^{(D)}_{\ell, \,{\bf k}}(r)\right]
\end{equation}
where ${\mathcal A}^{(3)}_{\ell}(\theta)=(2 \ell +1)  {\mathrm P}_{\ell}(\cos \theta)$ in terms of the Legendre polynomial ${\mathrm P}_{\ell}(x)$ and ${\mathcal A}^{(2)}_{\ell}(\theta)= 2\cos{(\ell \theta)}$. In Eq. (\ref{eq2}) the sum over $\ell$ runs over odd integers, because the spinor associated with the $\uparrow \uparrow$ state is symmetric and hence the coordinate part of the wave function has to be antisymmetric under exchange. The spinor associated with the $\uparrow\downarrow$ state has instead no definite symmetry and hence the sum in Eq. (\ref{eq3}) runs over all 
integer values of $\ell$. Upon performing the angular average in Eq. (\ref{eq7}) one obtains
\begin{equation}\label{eq10}
g_{\uparrow \uparrow}(r)=\frac{2}{r^{D-1}}\,\sum_{\ell=1\,\,({\mathrm odd})}^\infty\,{\mathcal B}^{(D)}_{\ell}\,\langle |\Phi^{(D)}_{\ell, {\bf k}}(r)|^2 \rangle_{p(k)}\, 
\end{equation}
and
\begin{equation}\label{eq11}
g_{\uparrow \downarrow}(r)=\frac{1}{r^{D-1}}\left[\langle|\Phi^{(D)}_{0, {\bf k}}(r)|^2\rangle_{p(k)}+\sum_{\ell=1}^\infty\,{\mathcal B}^{(D)}_{\ell}\,\langle |\Phi^{(D)}_{\ell, {\bf k}}(r)|^2 \rangle_{p(k)} \right]
\end{equation}
where ${\mathcal B}^{(3)}_{\ell}=2\ell+1$ and ${\mathcal B}^{(2)}_{\ell}=2$. Finally, the Schr\"{o}dinger equation for the wave function $\Phi^{(D)}_{\ell, {\bf k}}(r)$ is
\begin{equation}\label{eq1}
\left\{-\frac{\hbar^2}{2 \mu} \frac{d^2}{d r^2} +\frac{1}{2 \mu\,r^2}\,\left[{\bf L}^2_D+\frac{\hbar^2}{4}\,(D-1)(D-3)\right]+V(r)\right\}\Phi^{(D)}_{\ell, {\bf k}}(r)=\frac{\hbar^2k^2}{2\mu}\, \Phi^{(D)}_{\ell, {\bf k}}(r)
\end{equation}
where $\mu=m/2$ is the reduced mass of the electron pair, ${\bf L}^2_3=\hbar^2\,\ell(\ell+1)$ with $\ell=0,1,2,...$ and ${\bf L}^2_2=\hbar^2\,\ell^2$ with $\ell=0,\pm 1,\pm 2,...$ (the negative values of $\ell$ are accounted for by the choice ${\mathcal B}^{(2)}_{\ell}=2$ in Eqs.~(\ref{eq10}) and (\ref{eq11})). In solving Eq.~(\ref{eq1}) we impose as a boundary condition that $\Phi^{(D)}_{\ell, {\bf k}}(r)$ tends asymptotically to the free solution ({\it i.e.} the one which is obtained by setting $V(r)=0$) except for a phase shift.

The form of Eq. (\ref{eq10}) ensures that the relation $g_{\uparrow\uparrow}(0)=0$ is satisfied, since all functions $\Phi^{(D)}_{\ell, {\bf k}}(r)$ vanish at the origin for $\ell\neq 0$. The cusp condition on $g_{\uparrow\downarrow}(r)$ reads $d\ln{g_{\uparrow\downarrow}(r)}/dr|_{r=0}=1/a_B$ in $D=3$ and $d\ln{g_{\uparrow\downarrow}(r)}/dr|_{r=0}=2/a_B$ in $D=2$: following the argument given by Kimball\cite{Kimball}, it will be satisfied if $V(r)$ tends to the 
bare Coulomb potential for $r\rightarrow 0$. It is also easily seen\cite{gp} that the Hartree-Fock results for $g_{\uparrow\uparrow}(r)$ and 
$g_{\uparrow\downarrow}(r)$ are recovered if $V(r)$ is set to zero in Eq. (\ref{eq1}). 
Given a general scattering potential $V(r)$, there is no guarantee that the charge neutrality condition,
\begin{equation}\label{cn}
n\,\int d^D{\bf r}\,\left[g(r)-1\right]=-1~,
\end{equation}
is satisfied. We have numerical evidence that Eq.~(\ref{cn}) is fulfilled for our choice of $V(r)$ (see subsection A). The accuracy with which this happens depends in practice on the numerical solution of Eq.~(\ref{eq1}) and thus on the number of angular momentum states that are included in Eqs.~(\ref{eq2}) and (\ref{eq3}).~\cite{footnote} 

Before proceeding to present our choice for the potential $V(r)$ in Eq.~(\ref{eq1}), we report the expressions for the probability function $p(k)$ introduced in Eq.~(\ref{pofk}). Using the momentum distribution $n(k)=\theta(k_F-k)$ for the free Fermi gas, one obtains
\begin{equation}\label{eq8}
p_0(k)= 24\frac{k^2}{k^3_F}-36\frac{k^3}{k^4_F}+12\frac{k^5}{k^6_F}
\end{equation}
in $D=3$~\cite{gp} and
\begin{equation}\label{eq9}
p_0(k)= \frac{16\,k}{\pi\,k^2_F}\,\left[\arccos{\left(\frac{k}{k_{F}}\right)}-\frac{k}{k_F}\,\sqrt{1-\frac{k^2}{k^2_F}}
\, \right]
\end{equation}
in $D=2$.~\cite{perdew}. We have also evaluated $p(k)$ for interacting electrons using QMC data on the momentum distribution from the work of Ortiz and Ballone~\cite{Ortiz} in $3D$
 and of Conti~\cite{Conti} in $2D$. However, this led to only small changes in the pair distribution functions reported in Sect.~\ref{sec3} with the help of Eqs.~(\ref{eq8}) and (\ref{eq9}).    

\subsection{Hartree potential and self-consistency}

Overhauser's proposal\cite{Overhauser2} for calculating $g(0)$ in the $3D$ case was to approximate $V(r)$ by the 
electrical potential of a model consisting of an electron at the origin and a neutralizing sphere of uniformly
distributed charge with radius $r_s a_B=(4\pi n/3)^{-1/3}$. Gauss's law ensures that $V(r)$ vanishes outside the sphere,
and an approximate solution of the scattering problem could be obtained by an iterative procedure. In the work of 
Gori-Giorgi and Perdew\cite{gp} the same model for $V(r)$ was adopted to evaluate $g(r)$, but a full solution of the Schr\"odinger
equation (\ref{eq1}) was achieved. In the $2D$ case with $e^2/r$ interaction, on the other hand, the potential outside
a uniformly charged disk of radius $r_s a_B=(\pi n)^{-1/2}$ with an electron at its center does not vanish, since
the electrical force field extends outside the plane in which the electrons are moving. A more refined model is therefore
necessary.\cite{Polini}

Here we approximate $V(r)$ in Eq. (\ref{eq1}) by the Hartree potential due to the whole distribution of electrical 
charge and evaluate it with the help of Poisson's equation. More precisely, in the $3D$ case $V(r)$ is taken to satisfy
the equation
\begin{equation}\label{P$3D$}
\Delta_{r}V(r)=-4\pi e^2[\delta(r)+ n(g(r)-1)]\, ,
\end{equation}
where $\Delta_{r}$ is the radial Laplace operator. The appropriate Poisson equation for the Hartree potential
$V_{\mathrm H}(r,z)$ in the $2D$ case, with $r$ the radial distance in the electron plane and $z$ the vertical distance
from the plane, is
\begin{equation}\label{P$2D$}
(\Delta_{r}+\frac{d^2}{dz^2})V_{\mathrm H}(r,z)=-4\pi e^2[\delta(r)+ n(g(r)-1)]\delta(z)
\end{equation}
and what is needed is $V(r)=V_{\mathrm H}(r,0)$. Evidently, by solving Eq. (\ref{eq1}) in conjunction with 
Eq. (\ref{P$3D$}) or Eq. (\ref{P$2D$}) we obtain a self-consistent determination of the effective potential and of the
radial distribution function.

The solution of Eq. (\ref{P$3D$}) and Eq. (\ref{P$2D$}) is easily obtained by introducing Fourier transforms. We define
the structure factor $S(k)$ through the relation
\begin{equation}
S(k)=1+n\int d^{D} {\bf r}\, [g(r)-1]\, \exp{(-i {\bf k}\cdot{\bf r})}\, .
\end{equation}
It is then easily seen that the Fourier transform of $V(r)$ (${\widetilde V}(k)$, say) is given by 
\begin{equation}\label{vk}
{\widetilde V}(k)=v(k)\,S(k)
\end{equation}
where $v(k)=4 \pi e^2/k^2$ in $D=3$ and $v(k)=2 \pi e^2/k$ in $D=2$. To prove Eq. (\ref{vk}) in the $2D$ case, we
notice that the Fourier transform of $V_{\mathrm H}(r,z)$ from Eq. (\ref{P$2D$}) is 
\begin{equation}
{\widetilde V}_{\mathrm H}(k,k_z)=\frac{4 \pi e^2}{k^2+k^2_z}\,S(k)
\end{equation}
so that
\begin{equation}
{\widetilde V}_{\mathrm H}(k,z)\equiv \int\frac{dk_z}{2\pi}\,e^{ik_z\,z}\,{\widetilde V}_{\mathrm H}(k,k_z)=\frac{2 \pi e^2}{k}\,e^{-k|z|}\,S(k)\, ,
\end{equation}
and ${\widetilde V}(k)={\widetilde V}_{\mathrm H}(k,z=0)=2\pi e^2S(k)/k$.

Let us examine the asymptotic behaviours of the effective potential ${\widetilde V}(k)$ given in Eq.~(\ref{vk}). At large momenta the structure factor $S(k)$ tends to unity, so that ${\widetilde V}(k) \rightarrow v(k)$ and $V(r)$ tends to the bare Coulomb potential in the limit $r \rightarrow 0$. As already remarked, this property ensures that the cusp condition on $g(r)$ is satisfied. In the limit $k \rightarrow 0$, on the other hand, charge neutrality ensures the validity of the plasmon sum rule, which may be written in the form
\begin{equation}
\lim_{k \rightarrow 0} S(k)= \varepsilon_k/ \hbar \omega_{\rm pl}
\end{equation}
where $\varepsilon_k= \hbar^2 k^2/2m$ and $\omega_{\rm pl}$ is the leading term in the plasmon dispersion relation, given by $\omega_{\rm pl}=(4 \pi n e^2/m)^{1/2}$ in $3D$ and by $\omega_{\rm pl}=(2 \pi n e^2 k /m)^{1/2}$ in $2D$. Therefore,
\begin{equation}
\lim_{k \rightarrow 0} {\widetilde V}(k)= \hbar \omega_{\rm pl}/2n
\end{equation} 
in both dimensionalities. That is, the Fourier transform of our choice for the effective scattering potential tends in the long-wavelength limit to a constant in $D=3$ and to zero with a $k^{1/2}$ law in $D=2$.

\subsection{s-wave scattering length}

We complete this discussion by showing how the s-wave scattering length $a_{\rm sc}(r_s)$ can be evaluated from the numerical solution of the electron-electron scattering problem. The s-wave phase shift $\delta_{0}(k)$ is introduced through the large-distance behavior of the two-particle scattering state with $\ell=0$ at fixed momentum $k$,
\begin{equation}\label{eq20}
\Phi^{(D)}_{0, {\bf k}}(r) \sim \cos{[kr-(D-1)\frac{\pi}{4}+\delta_{0}(k)]}\, .
\end{equation}
From $\delta_0(k)$ the scattering length is obtained by the requirement that the wave function outside the range of the potential should vanish at $r=a_{\rm sc}$. In $D=3$ this yields from Eq.~(\ref{eq20}) the well-known relation
\begin{equation}
a_{\rm sc}(r_s)= -\lim_{k \rightarrow 0} \,\frac{\delta_0(k)}{k}.
\end{equation}
A simple analytical expression for $a_{\rm sc}(r_s)$ in the $3D$ electron gas is available from the work of Overhauser~\cite{Overhauser2},
\begin{equation}\label{$3D$sl}
a_{\rm sc}(r_s)=r_s a_{B}\,\frac{r_s/10}{1+3r_s/8}~.
\end{equation}

The introduction of the concepts of effective range and scattering length is much more delicate in $D=2$. In the work of Verhaar {\it et al}.~\cite{$2D$} the H-H atomic scattering problem was analyzed in detail. The most appropriate form of the outer wave function in $2D$ is
\begin{equation}
\Phi^{(2)}_{0, {\bf k}}(r) \sim \exp{[i \delta_0(k)]}\,\cos{[\delta_0(k)]}\{{\rm J}_0(kr)-\tan{[\delta_0(k)]}\,{\rm N}_0(kr)\}~,
\end{equation}
in terms of the Bessel functions ${\rm J}_0(x)$ and ${\rm N}_0(x)$. The relation between scattering length and phase shift thus is $\tan{[\delta_0(k)]}={\rm J}_0(ka_{\rm sc})/{\rm N}_0(ka_{\rm sc})$, taking at low energy a form which is the same as for the scattering of two hard spheres of radius $a_{\rm sc}$,
\begin{equation}\label{eq24}
\cot{[\delta_0(k)]}=(2/\pi)\,\left[\gamma +\ln{(ka_{\rm sc}/2)}\right] +{\rm o}(k^2)
\end{equation}
where $\gamma$ is the Euler's constant, $\gamma=0.577215665\cdots$ . In Sect.~\ref{sec3} we determine the scattering length for the $2D$ electron gas by fitting the expression in Eq.~(\ref{eq24}) to the phase shift obtained from the asymptotic behavior of the two-electron wave function as a function of $k$ at low energy. An analytical expression for $a_{\rm sc}$ in the strong-coupling limit has been given by Polini {\it et al}.~\cite{Polini}.

\section{Numerical results}\label{sec3}

\begin{figure}
\centerline{\mbox{\psfig{figure=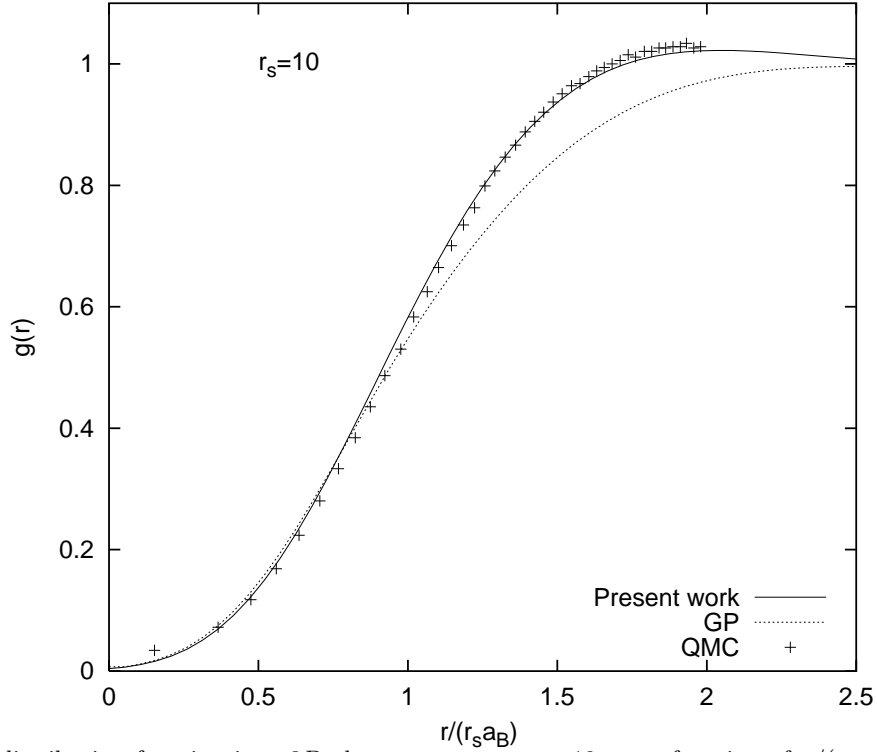, angle =0, width = 12 cm}}}
\caption{The pair distribution function in a $3D$ electron gas at $r_s=10$, as a function of $r/(r_s a_B)$. The results of the self-consistent Hartree approximation (full line) are compared with QMC data (crosses, from Ortiz {\it et al}.~\cite{Ortiz}) and with the results of calculations by Gori-Giorgi and Perdew~\cite{gp} (dotted line).}
\label{Fig1}
\end{figure}

\begin{figure}
\centerline{\mbox{\psfig{figure=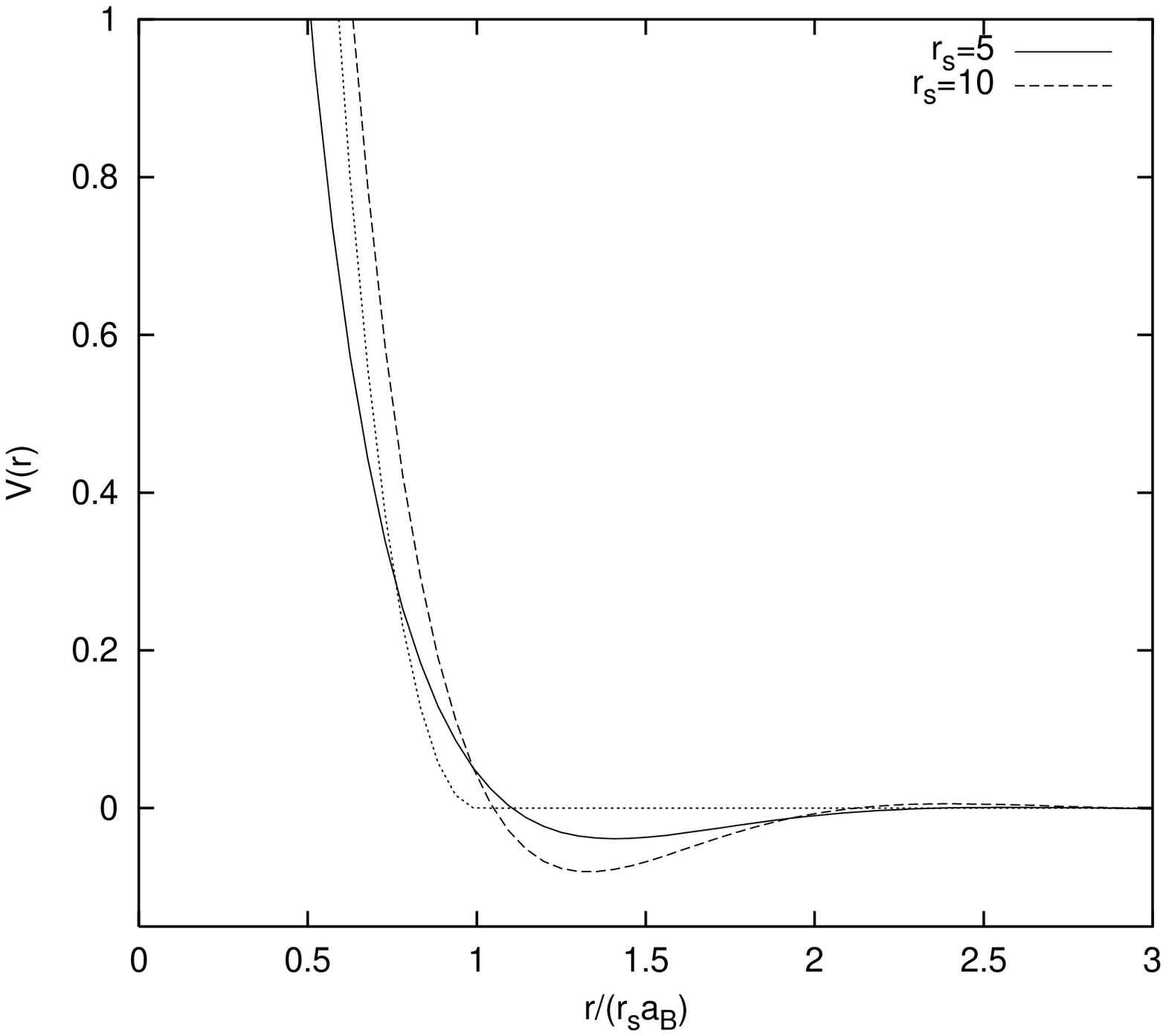, angle =0, width =9 cm}\psfig{figure=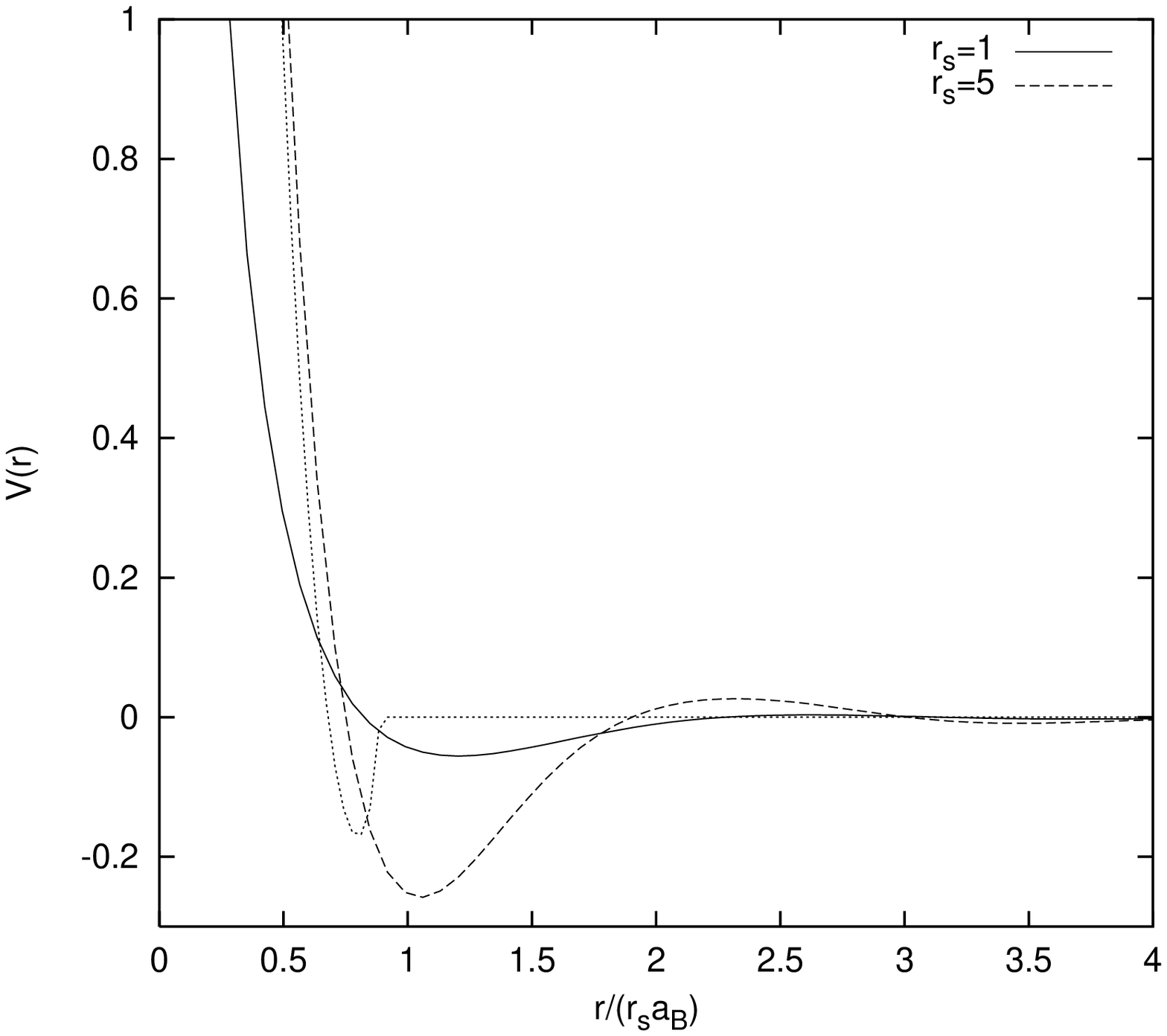, angle =0, width =9 cm}}}
\caption{Self-consistent scattering potential $V(r)$ in the Hartree approximation (in units of $\hbar^2 k^2_F/m$), as a function of $r/(r_s a_B)$ in $3D$ (left) and in $2D$ (right). The Overhauser potential in 3D\cite{Overhauser2} at $r_s=10$ and the result of the work of Polini {\it et al}.\cite{Polini}  in 2D at $r_s=5$ are reported for comparison (dotted lines).}
\label{Fig2}
\end{figure}

\begin{figure}
\centerline{\mbox{\psfig{figure=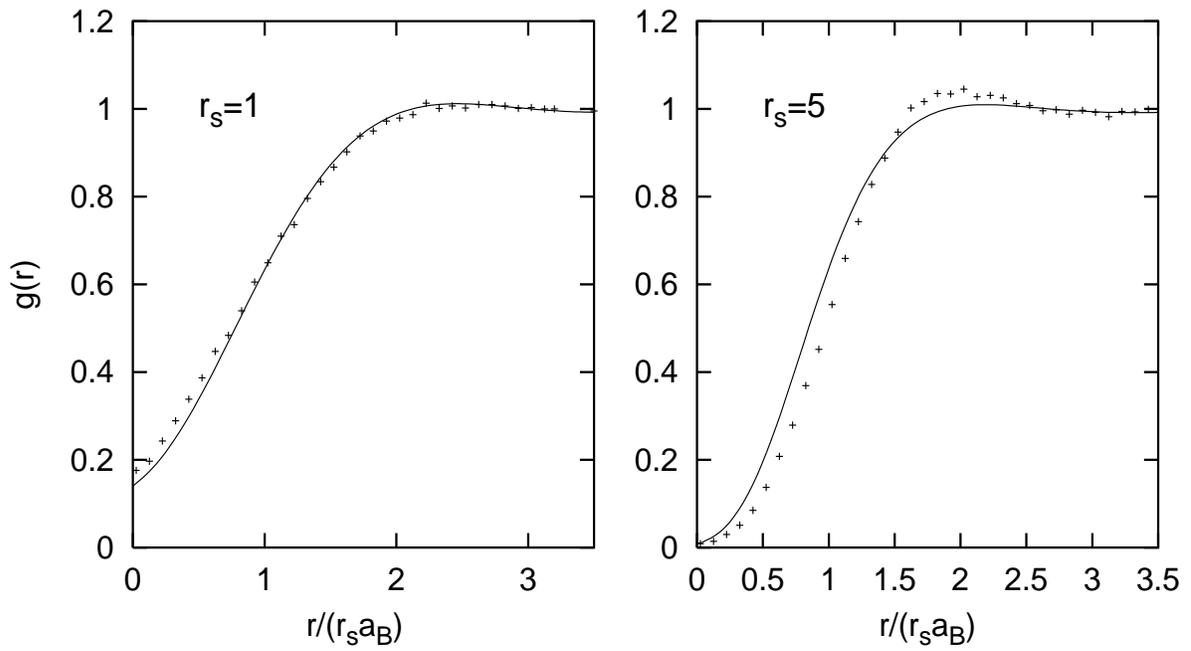, angle =0, width = 15 cm, height=22 cm}}}
\caption{The pair distribution function in a $2D$ electron gas at $r_s=1$ and $r_s=5$, as a function of $r/(r_s a_B)$. The results of the self-consistent Hartree approximation (full line) are compared with QMC data (crosses, from Tanatar and Ceperley~\cite{Tanatar}).}
\label{Fig3}
\end{figure}

\begin{figure}
\centerline{\mbox{\psfig{figure=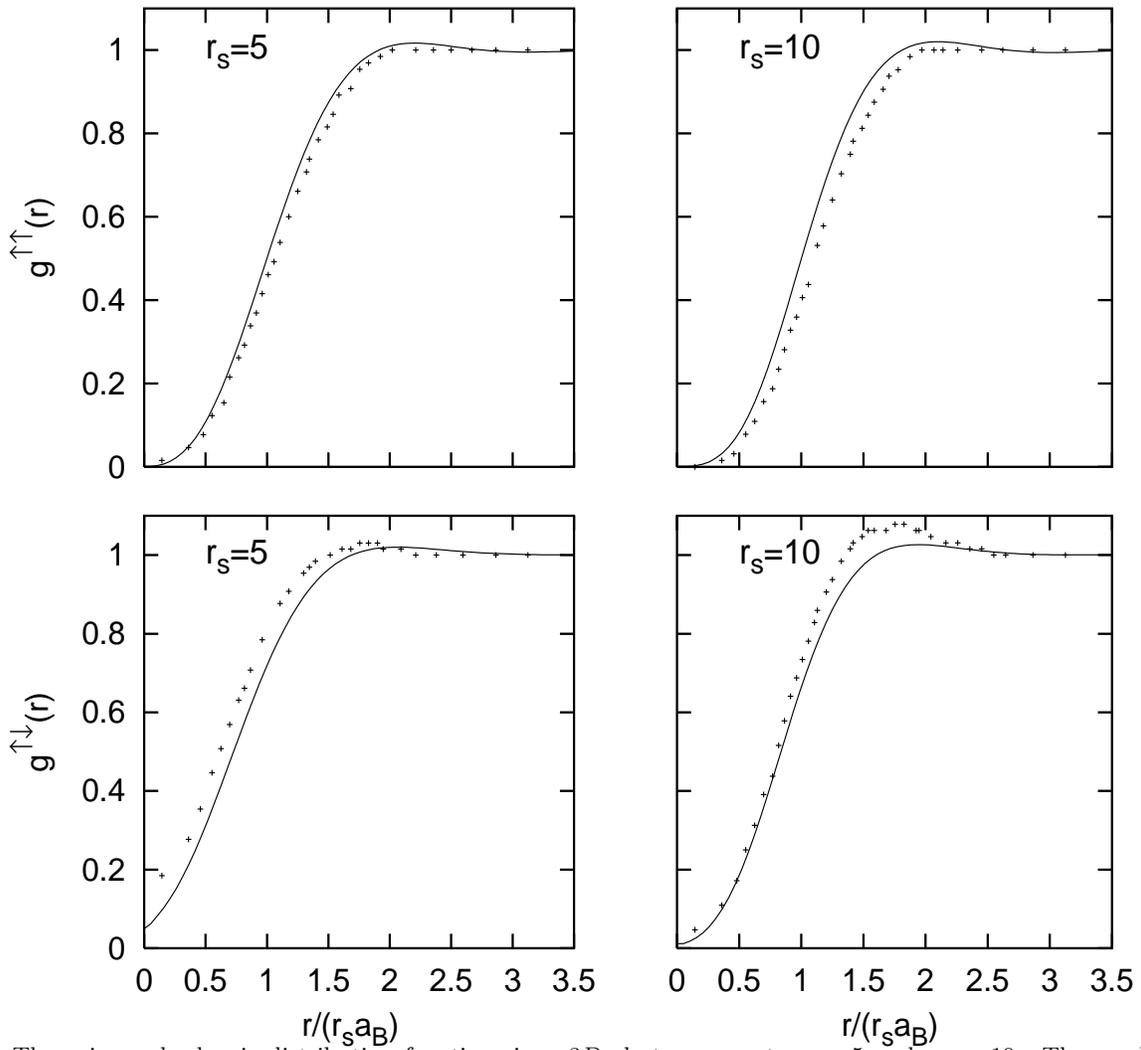, angle =0, width = 15 cm, height=22 cm}}}
\caption{The spin-resolved pair distribution functions in a $3D$ electron gas at $r_s=5$ and $r_s=10$. The results of the self-consistent Hartree approximation (full lines) are compared with the QMC data (crosses, from Gori-Giorgi {\it et al}.~\cite{Ortiz}).}
\label{Fig4}
\end{figure}

\begin{figure}
\centerline{\mbox{\psfig{figure=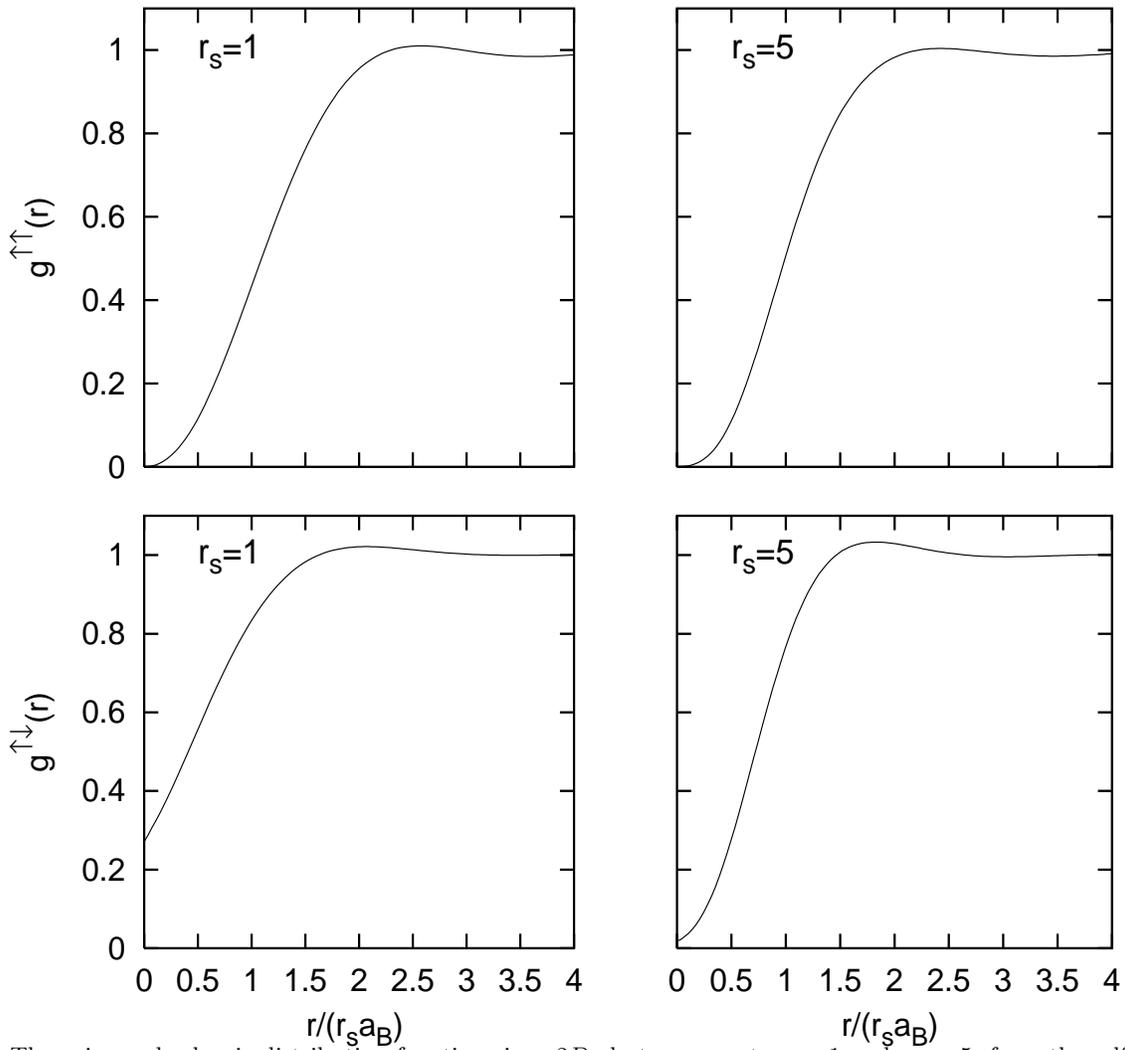, angle =0, width = 15 cm, height=22 cm}}}
\caption{The spin-resolved pair distribution functions in a $2D$ electron gas at $r_s=1$ and $r_s=5$, from the self-consistent Hartree approximation.}
\label{Fig5}
\end{figure}

We report in Figures \ref{Fig1} to \ref{Fig5} our numerical results for the pair distribution functions and for the self-consistent scattering potential. Starting with the $3D$ system, Figure \ref{Fig1} shows the spin-averaged $g(r)$ at $r_s=10$, in comparison with the QMC data reported by Ortiz {\it et al}..\cite{Ortiz} The results obtained by Gori-Giorgi and Perdew\cite{gp} within the same theoretical scheme, but with Overhauser's choice for the effective potential $V(r)$, are also shown in Figure \ref{Fig1}. It is seen from the Figure that both choices of $V(r)$ yield excellent agreement with the QMC data for the short-range part of the electron-electron correlations, up to $r/(r_s a_B)\approx 1$. However, the self-consistent calculation based on the use of
the Hartree potential becomes definitely superior at large distance, where (at this intermediate value of the coupling
strength) it continues to be in excellent agreement with the data. The Hartree potential at self-consistency is shown
in Figure \ref{Fig2}, for both the $3D$ and the $2D$ case. We may also remark that the cusp condition is satisfied by our
numerical results in both cases.

With further increase in the coupling strength the pair distribution function from the QMC work starts developing 
a first-neighbor peak, and this behavior is not reproduced quantitatively by the theory. This is illustrated in
Figure \ref{Fig3} for the electron gas in $D=2$. As is well known, the reduction in dimensionality enhances
the role of the electron-electron correlations: this is also clear from comparing the scattering potentials
in the two panels in Figure \ref{Fig2}. While in $2D$ the self-consistent theory remains quite accurate at moderate
coupling strength as is shown by the comparison with the QMC data of Tanatar and Ceperley\cite{Tanatar} at $r_s=1$ in
the left-hand panel in Figure \ref{Fig3}, quantitative differences from the QMC data are seen to arise at $r_s=5$ 
(right-hand panel in Figure \ref{Fig3}). 

The other aspect of the theory that needs testing concerns the quality of its predictions in regard to the
spin-resolved pair distribution functions. This point is examined in Figure \ref{Fig4} for the $3D$ system
at $r_s=5$ and $10$, using the QMC data of Ortiz {\it et al.} as analyzed by Gori-Giorgi {\it et al.}.\cite{Ortiz}
The discrepancies between theory and simulation are reasonably small at these values of the coupling strength. It is
evident from the Figure that these discrepancies largely cancel out in taking the spin avarage, but will be magnified when one calculates the difference distribution function $g_d(r)$. Similar theoretical results are shown in Figure \ref{Fig5} for the $2D$ system.

\begin{figure}
\centerline{\mbox{\psfig{figure=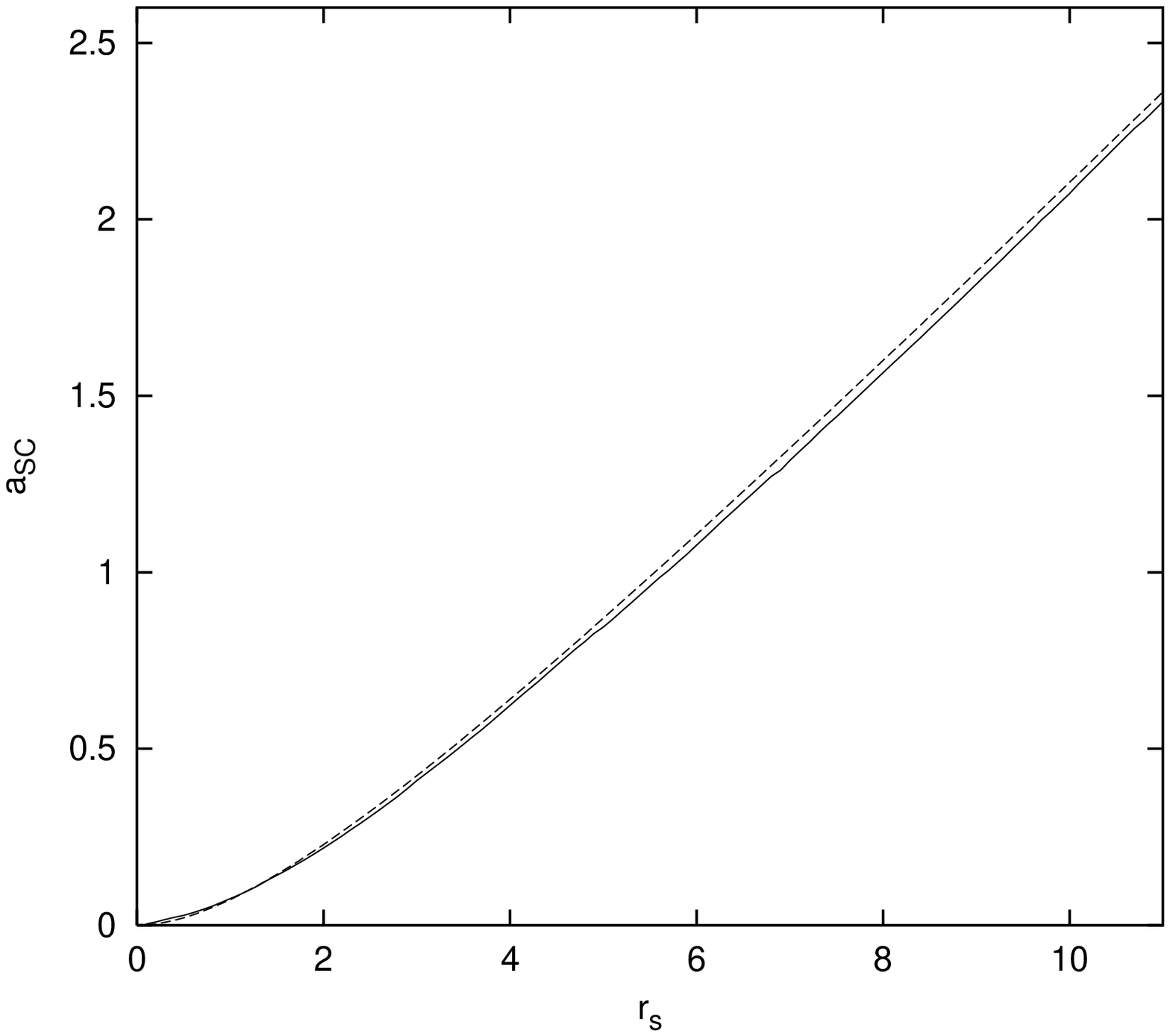, angle =0, width =9 cm}\psfig{figure=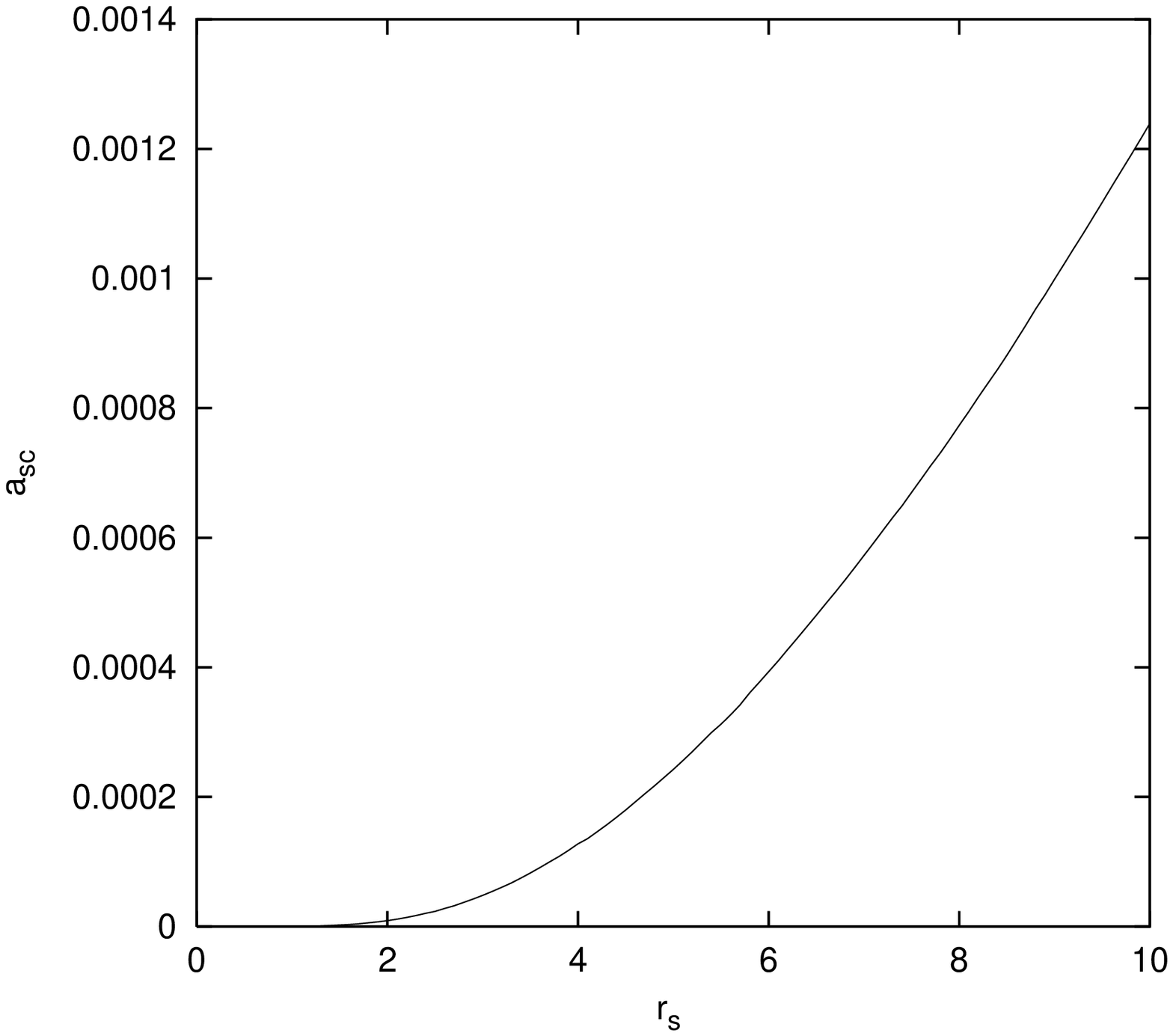, angle =0, width =9 cm}}} 
\caption{The s-wave scattering length as a function of $r_s$, in units of the Bohr radius. Left panel: the result of the self-consistent Hartree approximation for the $3D$ electron gas (full line) is compared with the analytical result of Overhauser~\cite{Overhauser2} in Eq.~(\ref{$3D$sl}) (broken line). Right panel: results from the self-consistent Hartree approximation for the $2D$ electron gas.}
\label{Fig6}
\end{figure}

\begin{figure}
\centerline{\mbox{\psfig{figure=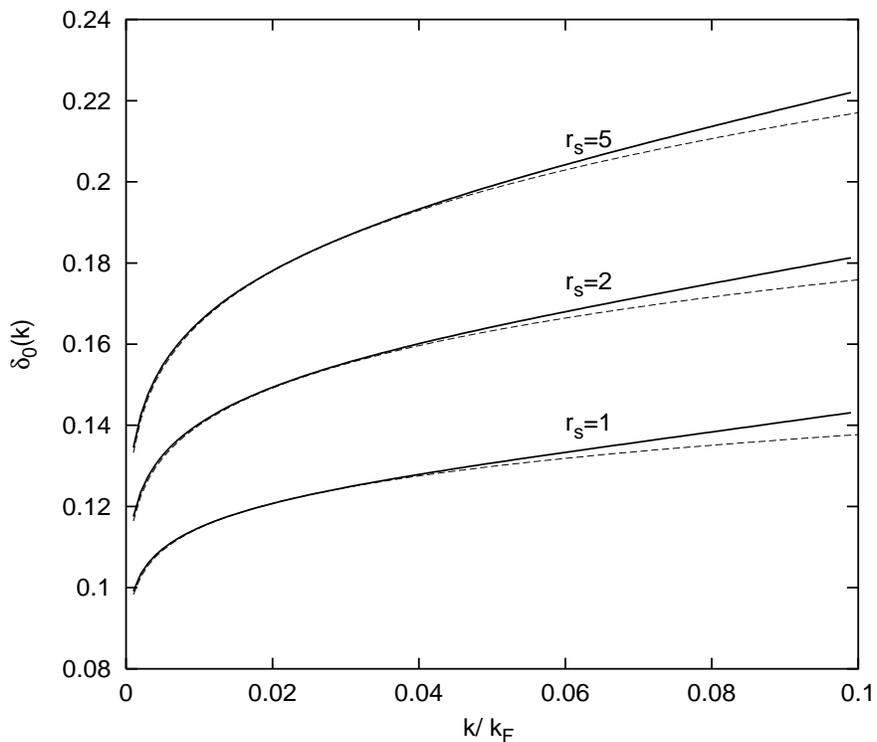, angle =0, width = 12 cm, height=10 cm}}}
\caption{The s-wave phase shift in the $2D$ electron gas at various values of $r_s$, as a function of reduced momentum $k/k_F$ at low momenta. Full lines: from the numerical solution of the self-consistent Hartree approximation; broken
lines: from Eq.~(\ref{eq24}) with the values of the scattering length shown in the right-hand panel in Figure \ref{Fig6}.}
\label{Fig7}
\end{figure}

Finally, in Figures \ref{Fig6} and \ref{Fig7} we report our results for the s-wave scattering length as a function
of coupling strength $r_s$. For the $3D$ system the simple analytical formula obtained by Overhauser\cite{Overhauser2} and reported in Eq. (\ref{$3D$sl}) is seen in the left panel in Figure \ref{Fig6} to give a very good account of our results. Results for the $2D$ electron gas at moderate coupling strength are shown in the right-hand panel in Figure \ref{Fig6}, while Figure \ref{Fig7} shows how they have been obtained by fitting the expression (\ref{eq24}) to our numerical results for the phase shift in s-wave as a function of momentum at low momenta. It may be remarked that the magnitude of $\delta_0(k)$ in the present electron-electron scattering in $2D$ is smaller than that in the H-H scattering problem studied by Verhaar {\it et al}.~\cite{$2D$} by a factor of about $2$. This yields, however, huge differences in the magnitude of the scattering length. Our results in the right-hand panel in Figure \ref{Fig6} should thus be regarded as being very sensitive to the details of the theory and hence as having limited quantitative value.

\section{Extension to inhomogeneous systems}\label{sec4}
In this section we briefly indicate how the approach that we have presented in Sect.~\ref{sec2} could be extended to deal with the pair distribution function in an inhomogeneous electron system, subject to an external scalar potential $V_{\rm ext}({\bf r})$. 

The electron-electron correlations are described in such a system by an inhomogeneous pair distribution function, $g({\bf r}, {\bf r}')$ say. The exchange and correlation energy functional is given by
\begin{equation}\label{25}
E_{\mathrm xc}=\frac{1}{2}\,\int d^{D}{\bf r}\int d^{D}{\bf r'}\, n({\bf r}) n({\bf r'})\,[{\bar g}({\bf r}, {\bf r'})-1]\,v(|{\bf r}-{\bf r}'|)\, ,
\end{equation}
where $n({\bf r})$ is the inhomogeneous electron density and ${\bar g}({\bf r}, {\bf r'})$ is obtained from 
$g({\bf r}, {\bf r'})$ by an integration over the coupling strength at fixed $n({\bf r})$ (see for instance the book by Dreizler and Gross~\cite{Barth}). The calculation of $g({\bf r}, {\bf r'})$ by means of a two-electron scattering problem remains in this case a genuine two-body problem. It requires for each spin state solution of the equation
\begin{equation}
\left[-\frac{\hbar^2}{2 m}\,( \Delta_{{\bf r}_1} +\Delta_{{\bf r}_2})+V_{\mathrm ext}({\bf r}_1, {\bf r}_2)+
V({\bf r}_1, {\bf r}_2)\right]
\Phi_{\displaystyle \epsilon_{12}}({\bf r}_1, {\bf r}_2)=\epsilon_{12}\Phi_{\displaystyle \epsilon_{12}}({\bf r}_1, {\bf r}_2)\, ,
\end{equation}
where $\Delta_{{\bf r}}$ is the Laplace operator, 
$V_{\mathrm ext}({\bf r}_1, {\bf r}_2)=V_{\mathrm ext}({\bf r}_1)+V_{\mathrm ext}({\bf r}_2)$, and $V({\bf r}_1, {\bf r}_2)$ is the effective electron-electron potential. The Hartree approximation on the effective potential, leading to
\begin{equation}\label{eff_in}
V({\bf r}_1, {\bf r}_2)=v(|{\bf r}_1-{\bf r}_2|)+ \int d^{D}{\bf r'}\,[n({\bf r}')\,g({\bf r}_1, {\bf r}')-n_b({\bf r}')]\, 
v(|{\bf r}'-{\bf r}_2|)
\end{equation}
provides an approximate self-consistent closure of the problem transcending the usual LDA or other approaches that appeal to the exchange-correlation hole of the homogeneous electron gas. In the previous equation (\ref{eff_in}), $n_b({\bf r}')$ is the density of the background.

Finally, a relation between $g({\bf r}, {\bf r'})$ and the two-electron scattering states  $\Phi_{\displaystyle \epsilon_{12}}({\bf r}, {\bf r}')$ is needed. This relation can in general be written in the following form:
\begin{equation}\label{summation}
g_{\sigma\sigma'}({\bf r}, {\bf r}')=\sum_{\epsilon_{12}, \,{\rm occup.}}\Gamma^{\sigma\sigma'}_{\epsilon_{12}}\,|\Phi_{\displaystyle \epsilon_{12}}({\bf r}, {\bf r}')|^2~,
\end{equation}
where the sum runs over all the occupied levels $\epsilon_{12}$. The appropriate degeneracy factor  $\Gamma^{\sigma\sigma'}_{\epsilon_{12}}$ for the eigenvalue $\epsilon_{12}$ is zero in the case $\sigma=\sigma'$, if $\Phi_{\displaystyle \epsilon_{12}}({\bf r}, {\bf r}')$ is symmetric under the exchange ${\bf r} \leftrightarrow {\bf r}'$.

For building the exchange and correlation energy functional in Eq.~(\ref{25}), one needs to calculate the pair distribution function at each given coupling strength $\lambda$ by repeating the procedure outlined above with $v(|{\bf r}_1-{\bf r}_2|)=\lambda/|{\bf r}_1-{\bf r}_2|$. The density profile $n({\bf r})$ is needed at full coupling strength and requires to be obtained by a parallel DFT procedure. 
The pair distribution function at full strength is, in itself, a very interesting quantity.

As an example of application of such scheme, we would like to mention the problem of a finite number of electrons confined in a quantum dot\cite{kouwenhoven}. In this case the eigenvalues $\epsilon_{12}$ are discrete and the summation procedure reported in Eq.~(\ref{summation}) corresponds to filling the lowest energy states with all the available electrons.

Another problem of interest is represented by a system of electrons confined in a quantum well, in cases where the electron dynamics is important also in the growth direction so that the confinement cannot be handled by a simple reduction to $2D$. Compared to the previous example, the summation procedure is in this case more involved. The difficulty comes from the fact that the motion in the transverse direction is free, and this implies that for each subband in the growth direction there is a dispersion in the transverse direction associated with the in-plane momentum. Examples of such physical systems, which are in a sense intermediate between $3D$ and $2D$, have been discussed for instance by Ullrich and Vignale~\cite{Ullrich} and by Luin {\it et al.}~\cite{Luin}. 

More generally, the possibility of testing theories of the exchange-correlation hole and of the exchange-correlation energy density in strongly inhomogenous electronic systems is opening up through novel applications of variational QMC methods~\cite{Needs}.
\section{Summary}
In this work we have proposed an extension of Overhauser's model for the electron-electron correlations in the $3D$ electron gas on the basis of a self-consistent Hartree approximation for the electron-electron scattering potential. We have confirmed that the model is quite accurately describing the short-range part of the exchange-correlation hole, as already demonstrated by Gori-Giorgi and Perdew~\cite{gp}, and shown that it can be usefully extended to cover the full range of interelectronic separation over an appreciable range of values of the coupling strength. As already noted in Sect.~\ref{sec3}, the accuracy of the present Hartree approximation is limited to the range of values of $r_s$ below the development of a first-neighbor peak in the pair distribution function. We have also shown that the original proposal of Overhauser yields a very accurate analytical formula for the electron-electron scattering length, and indicated how this approach could be extended to deal with the exchange-correlation hole in an inhomogeneous electron gas, leading perhaps to more accurate descriptions of the exchange-correlation energy functional.

We have examined the usefulness of this approach in describing the exchange-correlation hole in the $2D$ electron gas and the spin-resolved pair distribution functions in the $3D$ electron gas. As may be expected, the Hartree approximation is quantitatively useful in $2D$ over a more limited range of coupling strength and has more limited accuracy in regard to the splitting of the exchange-correlation hole into its parallel- and antiparallel-spin components. With regard to these spin-resolved pair functions, however, we feel that it is quite remarkable that a Hartree approximation should already work as well as it does in our calculations. We may hope that major changes will not be needed to explicitly include many-body exchange corrections in the electron-electron scattering potential.

We have also given some attention to the determination of the electron-electron scattering length in the $2D$ electron gas. As is generally true for $2D$ systems, the logarithmic dependence of the s-wave phase shift on $k a_{\rm sc}$ at very low energy makes this problem a very delicate one. We have shown that the proposal made by Verhaar {\it et al}.~\cite{$2D$} in the context of an atom-atom scattering process is just as useful in regard to electron-electron scattering.

We are currently examining how the Hartree scattering potential may be supplemented by explicit inclusion of corrections accounting for many-body exchange and correlations, with the main aims of (i) improving the quantitative account of the spin-resolved components of the exchange-correlation hole, and (ii) studying how a first-neighbor shell emerges with increasing coupling strength.
\acknowledgements
This work was partially supported by MIUR through the PRIN2001 program.

\newpage

\begin{references}
\bibitem{footnote}
In our calculations we have found that satisfying Eq.~(\ref{cn}) at the level of a percent may require as many as 40 angular-momentum states in $3D$ and even more in $2D$.
\end{references}
\end{document}